%% file: a1351_mnras_v4.tex
\documentclass[useAMS,usenatbib]{mn2e}

\usepackage{graphicx}

\newcommand{\mincir}{\raise -2.truept\hbox{\rlap{\hbox{$\sim$}}\raise5.truept
\hbox{$<$}\ }}
\newcommand{\magcir}{\raise -2.truept\hbox{\rlap{\hbox{$\sim$}}\raise5.truept
\hbox{$>$}\ }}
\newcommand{\siml}{\raise -2.truept\hbox{\rlap{\hbox{$\sim$}}\raise5.truept
\hbox{$<$}\ }}
\newcommand{\simg}{\raise -2.truept\hbox{\rlap{\hbox{$\sim$}}\raise5.truept
\hbox{$>$}\ }}
\newcommand{\be}{\begin{equation}}
\newcommand{\ee}{\end{equation}}
\newcommand{\ba}{\begin{eqnarray}}
\newcommand{\ea}{\end{eqnarray}}
\newcommand {\kpc} {$h_{70}^{-1}$ kpc $\;$}

\newcommand {\hh} {$h_{70}^{-1}$ Mpc}
\newcommand {\hhh} {\;h_{70}^{-1} \mathrm{Mpc}}
\newcommand {\ks} {km~s$^{-1} \;$}
\newcommand {\kss} {km~s$^{-1}$}

\newcommand {\mquaa} {$\times 10^{14}\;h_{70}^{-1}\;M_{\odot}$}
\newcommand {\mqui} {$\times 10^{15}\;h_{70}^{-1}\;M_{\odot} \;$}
\newcommand {\mquii} {$\times 10^{15}\;h_{70}^{-1}\;M_{\odot}$ }

\newcommand {\mll} {$h_{70}\;M_{\odot}/L_{\odot}$}

\newcommand{\degree}{\ensuremath{\mathrm{^\circ}}}
\newcommand{\arcm}{\ensuremath{\mathrm{^\prime}\;}}
\newcommand{\arcs}{\ensuremath{\arcmm\hskip -0.1em\arcmm \;}}
\newcommand{\arcmm}{\ensuremath{\mathrm{^\prime}}}

\newcommand{\dotsec}{\,\rlap{\hbox{$\mathrm{^s}$}}{\hbox{$.$}}\,}

\title[]{The structure of Abell 1351: a bimodal galaxy cluster 
with peculiar diffuse radio emission}
\author[R. Barrena, M. Girardi, W. Boschin, S. De Grandi, M. Rossetti]{R. 
Barrena$^{1,2}$\thanks{E-mail: rbarrena@iac.es}, M. Girardi$^{3,4}$, W. Boschin$^{5}$,
S. De Grandi$^{6}$, M. Rossetti$^{7,8}$ \\
$^{1}$Instituto de Astrof\'{\i}sica de Canarias, C/V\'{\i}a L\'actea s/n, E-38205 La Laguna (Tenerife), Spain\\
$^{2}$Departamento de Astrof\'{\i}sica, Univ. de La Laguna, Av. del
     Astrof\'{\i}sico Francisco S\'anchez s/n, E-38205 La Laguna
     (Tenerife), Spain\\
$^{3}$Dipartimento di Fisica dell'Universit\`a degli Studi
     di Trieste - Sezione di Astronomia, via Tiepolo 11, I-34143
     Trieste, Italy\\ 
$^{4}$INAF - Osservatorio Astronomico di Trieste,
     via Tiepolo 11, I-34143 Trieste, Italy\\ 
$^{5}$Fundaci\'on Galileo Galilei - INAF (Telescopio Nazionale Galileo), 
     Rambla J. A. Fern\'andez P\'erez 7, E-38712 Bre\~na Baja (La Palma), Spain\\ 
$^{6}$INAF, Osservatorio Astronomico di Brera, via E. Bianchi 46, 23807, Merate (LC), Italy\\
$^{7}$Dipartimento di Fisica dell'Universit\`a degli Studi di Milano, via Celoria 16,
     I-20133, Milan, Italy\\
$^{8}$IASF-Milano, INAF, via Bassini 15, Milan 20133, Italy}

\begin{document}

\date{Accepted DATE. Received DATE; in original form DATE}

\pagerange{\pageref{firstpage}--\pageref{lastpage}} \pubyear{2012}

\maketitle

\label{firstpage}

\begin{abstract}
We aim to review the internal structure and dynamics of the 
Abell 1351 cluster, shown to host a radio halo with a quite irregular shape.
Our analysis is based on radial velocity data for 135 galaxies
obtained at the Telescopio Nazionale Galileo. We combine galaxy
velocities and positions to select 95 cluster galaxy members and
analyse the internal dynamics of the whole cluster. We also examine
X-ray data retrieved from {{\em Chandra}} and {{\em XMM}} archives.
We measure the cluster redshift, $\left<z\right>=0.325$, the
line-of-sight (LOS) velocity dispersion, $\sigma_{\rm V}\sim 1500$
\kss, and the X-ray temperature, $kT\sim$ 9 keV. From both X-ray and
optical data independently, we estimate a large cluster mass, in the
1--4 \mqui range. We attribute the extremely high value of
$\sigma_{\rm V}$ to the bimodality in the velocity distribution. We
find evidence of a significant velocity gradient and optical 3D
substructure. The X-ray analysis also shows many features in favour
of a complex cluster structure, probably supporting an ongoing merger of
substructures in Abell 1351. The observational scenario agrees with
the presence of two main subclusters in the northern region, each with
its brightest galaxy (BCG1 and BCG2), detected as the two most
important X-ray substructures with a rest-frame LOS velocity
difference of $\Delta V_{\rm rf}\sim 2500$ \kss and probably being in
large part aligned with the LOS.  We conclude that Abell 1351 is a
massive merging cluster. The details of the cluster structure allow us to
interpret the quite asymmetric radio halo as a `normal' halo plus a
southern relic, strongly supporting a previous suggestion based only 
on inspection of radio and preliminary X-ray data.

\end{abstract}


\section{Introduction}
\label{intro}

Very few galaxy clusters show powerful diffuse radio emission on a large
scale. In general, this emission may be classified as `radio haloes'
and `radio relics'. In the first case, the emission presents a
compact morphology and comes from central cluster regions, while
`relics' are located in the peripheral zones
\citep{fer08,gio02,fer12,ven11}. The synchrotron origin of this radio
emission demonstrates the presence of large-scale magnetic fields and
relativistic particles distributed throughout the whole
cluster. Today, cluster mergers are accepted as the most suitable
scenarios proposed to supply enough energy for accelerating electrons
to relativistic velocities and for magnetic field amplification.

In this framework, radio relics seem to be directly linked with merger
shocks \citep{hoe04,ens01,roe99,ens98}. On the other hand, the
remaining turbulence after cluster collision has been proposed as one
of the most important mechanisms to produce giant radio haloes
\citep{bru01,bru09}.  Nevertheless, the precise mechanism for
generating powerful radio haloes in clusters is still debated. In fact,
there are two main theoretical approaches to the problem:
re-acceleration vs.\ hadronic models (Brunetti et al. \citeyear{bru09}
and references therein).

Traditionally, the dynamical state of clusters with diffuse radio
emission is studied using X-ray observations. Moreover, all
statistical analyses on properties of radio emission are derived in
general from X-ray data \citep{ros11,cas10,buo02,sch01}, and consist basically of
temperature and luminosity \citep[see e.g.][and references
  therein]{gio02}. In fact, simulations based on turbulent
re-acceleration models fit the observed radio haloes reasonably well 
\citep{cas06}. Following the same line, \cite{gov01} also find very
good agreement between X-ray and radio emission when comparing,
point by point, their respective surface brightnesses.  Furthermore,
\citet{bas12} reports no evidence of bimodality in the radio-power--integrated SZ effect diagram whereas, on the contrary, \citet{bru07}
find a bimodal distribution in the radio-power--X-ray luminosity
plane. This controversy highlights the need to investigate the nature
of radio haloes using other research techniques in addition to X-ray
data.

The information on cluster members retrieved by radial velocities is
an excellent way to study the kinematics and dynamics of mergers in
clusters \citep{gir02}. This technique allows us to detect cluster
substructures by distinguishing pre- and post-merging groups, thus
making it possible to disentangle the dynamics of ongoing mergers and
remnants. Moreover, optical data complement and add information to
X-ray because, during a collision, the ICM and galactic component
respond on different timescales.  Numerical simulations by
\cite{roe97} clearly support this fact, hence the importance of
combining optical and X-ray data to study the interactions of
different cluster constituents. This is also the technique carried out
by the MUSIC (MUlti-Wavelength Sample of Interacting Clusters) project
\citep{maur11} to study merging scenarios.

In the context described above, we are conducting an intensive
observational and data analysis programme to study the internal dynamics
of clusters with diffuse radio emission by using member galaxies (DARC
-- Dynamical Analysis of Radio Clusters -- project, see Girardi et al.\ 
\citeyear{gir10conf}.\footnote{See also http://wwwuser.oats.inaf.it/girardi/darc/, 
the web site of the DARC project.})  As part of an ongoing programme to 
investigate  the presence in complex X-ray cluster systems of halo emission 
from the sub-clusters in the merger, we have carried out a detailed and
intensive study of the cluster Abell 1351 (hereafter A1351), also
known as MACS J1142.4+5831. A1351 is a rich, X-ray luminous cluster:
Abell richness class $=2$ \citep{abe89}; $L_\mathrm{X}$(0.1--2.4
keV) = 8.31$\times 10^{44} \ h_{50}^{-2}$ erg\ s$^{-1}$ \citep{boe00}.
\citet{boe00} report a mass distribution elongated in the NNE--SSW
direction (see also Dahle et al.\ \citeyear{dah02}), but the available
maps do not agree in detail. In fact, the primary peak of
\citet{hol09} corresponds to the secondary one of Dahle et
al.\ (\citeyear{dah02}) (cf.\  figure~3 -- top-left panel -- of Holjem et
al.\ with figure~17 -- top-right panel -- of Dahle et al.). The \textit{ROSAT} HRI
image also reveals a clear elongation, roughly in the north--south
direction \citep{gia09,gio09}, thus suggesting ongoing merger
activity \citep{all03}.

\begin{figure*}
\centering 
\includegraphics[width=18cm]{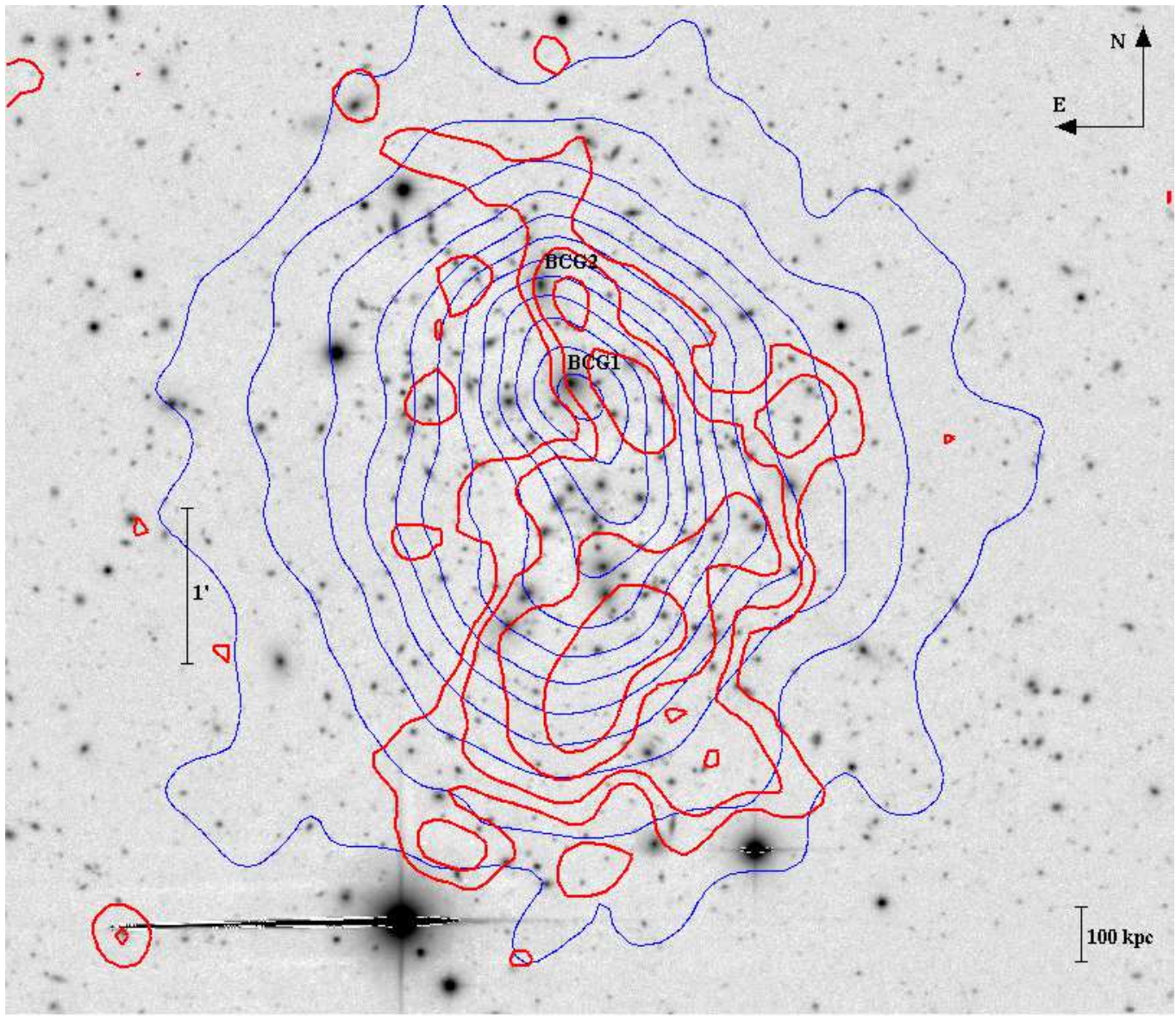}
\caption{INT $R$-band image of the cluster A1351 (north at the top
and east to the left) with, superimposed, the contour levels of the
\textit{Chandra} archival image ID~15136 (blue thin contours; photons in the
energy range 0.3--7 keV) and the contour levels of a VLA radio image
at 1.4 GHz (red thick contours, see Giacintucci et
al.\ \citeyear{gia09}). Labels highlight the positions of the two
brightest cluster galaxies.}
\label{figimage1}
\end{figure*}

As for the radio emission, \citet{owe99} reported the first evidence
of a diffuse, extended source in this cluster. The analysis of VLA
data archive at 1.4 GHz shows the presence of a centrally located,
Mpc-extended source classified as a giant radio halo with a power 
$P_{\rm 1.4GHz}=1.2$-$-1.3\times10^{25}\ h_{70}^{-2}$ W\ Hz$^{-1}$
\citep{gia09,gio09}. Despite the particular difficulty of the subtraction
of discrete sources present in radio maps, both analyses of
\citet{gia09} and \citet{gio09} show that the radio diffuse emission
is quite asymmetric with respect to the X-ray peak. In particular,
\citeauthor{gia09} suggest that the ridge in the southern peripheral
region of the halo is a relic projected on to the radio halo emission
(see their figure~2 and our Figure~\ref{figimage1}).  A1351 is also an
atypical cluster for its very large velocity dispersion of
$\sigma_v=1680_{-230}^{+340}$ \kss. This value and the published
cluster redshift ($z=0.3279$) are based on 17 cluster members
\citep{irg02}.

 We acquired new spectroscopic and photometric data on the 3.5 m
Telescopio Nazionale Galileo (TNG) and the 2.5 m Isaac Newton Telescope
(INT), respectively. The present work is mainly based on these optical
data. We also used optical data from the Sloan Digital Sky Survey,
Data Release 7 (hereafter SDSS-DR7). We also use X-ray data from {\it
  Chandra} and {\it XMM-Newton} public archive to study the
intra-cluster medium of A1351 and analyse its global morphological and
spectral properties.

This paper is organized as follows. We present optical data and the
cluster catalogue in Section~2. Our results on the cluster structure
based on optical and X-ray analyses are presented in Sects~3 and 4,
respectively. We discuss the cluster structure in Sect.~5 and present
our conclusions in Section~6. Unless otherwise stated, we indicate
errors at the 68\% confidence level (hereafter c.l.). Throughout this
paper, we use $H_0=70$ km s$^{-1}$ Mpc$^{-1}$ and $h_{70}=H_0/(70$ km
s$^{-1}$ Mpc$^{-1}$) in a flat cosmology with $\Omega_0=0.3$ and
$\Omega_{\Lambda}=0.7$. In the adopted cosmology, 1\arcm corresponds
to $\sim$282 \kpc at the cluster redshift.

\section{Optical observations}
\label{data}

\begin{figure*}
\centering 
\includegraphics[width=18cm]{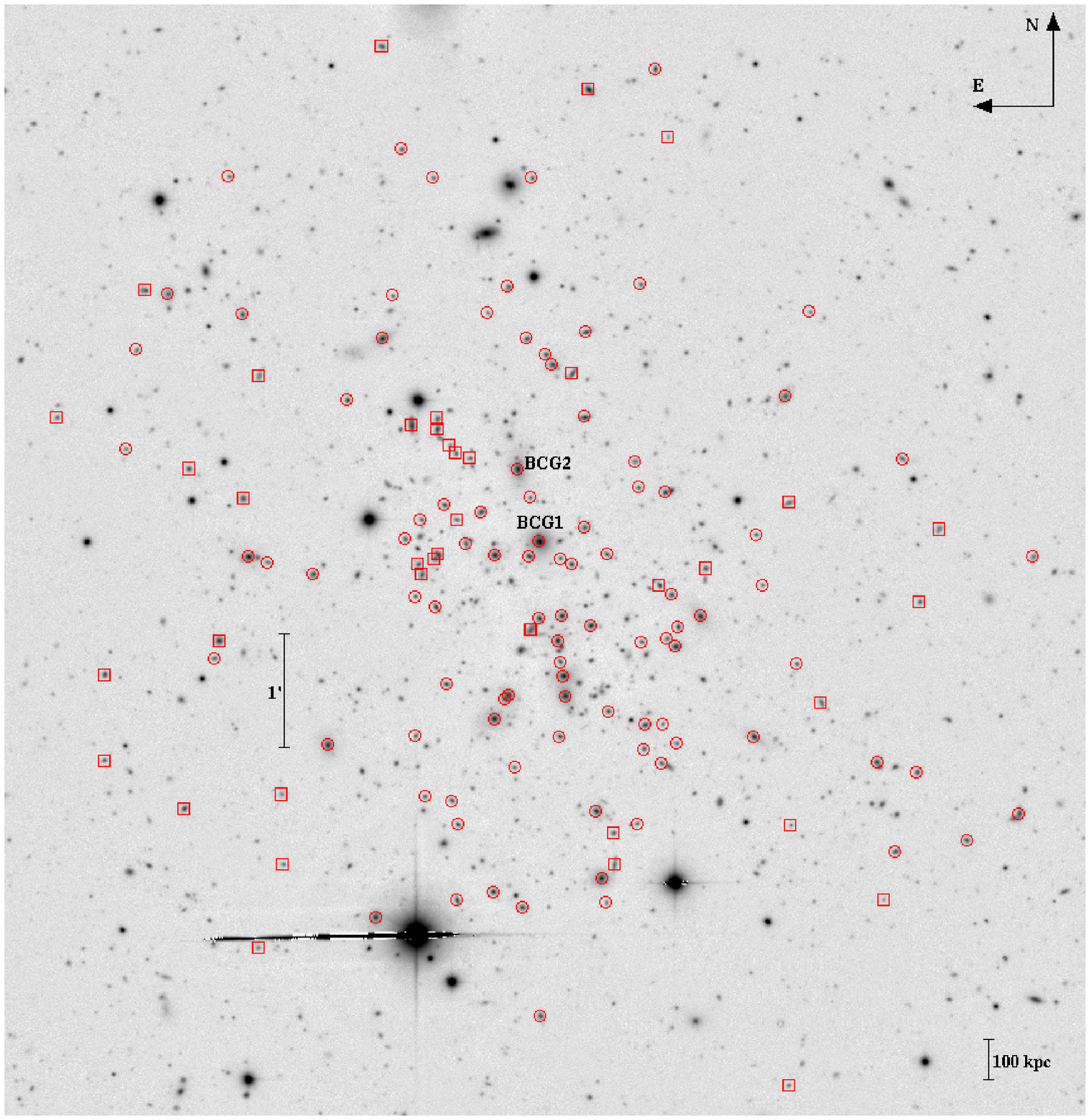}
\caption{Galaxy sample observed spectroscopically, overlapping the
  INT $R$-band image of the cluster A1351. Circles and squares
  indicate cluster members and non-members, respectively (see
  Table~\ref{catalogA1351}). Labels indicate the brightest galaxy
  members.}
\label{figottico1}
\end{figure*}

Multi-object spectroscopic (MOS) observations of A1351 were carried
out at the TNG on 2010 March 10. We used DOLORES/MOS with the
LR-B Grism 1, yielding a dispersion of 187 \AA/mm.  We used the
$2048\times2048$ pixel E2V CCD, with a pixel size of 13.5 $\mu$m. In
total, we observed four MOS masks including 143 slits. For each mask, the
exposure time was $3\times1800$ s.

Reduction of spectra and radial velocity computation were performed
using standard IRAF\footnote{IRAF is distributed by the National
  Optical Astronomy Observatories, which are operated by the
  Association of Universities for Research in Astronomy, Inc., under
  cooperative agreement with the National Science Foundation.} tasks
and the cross-correlation technique developed by Tonry \& Davis
(\citeyear{ton79}), as done with other clusters included in our DARC
sample (for a detailed description see, for example, \citealt{bos12}). Our
spectroscopic catalogue lists 129 galaxies in the field of A1351 (see
Table~\ref{catalogA1351}).

Comparing the velocity measurements for ten galaxies observed with
multiple masks, we corrected the velocity errors provided by the
cross-correlation technique by multiplying them for a factor 2.2 (see
discussion in, e.g., \citealt{gir11}, \citealt{bos04}). Taking into
account the above correction, the median value of the $cz$ errors is
77 \kss. Three galaxies have spectroscopic redshift in SDSS-DR7 in
good agreement with our values. We used magnitudes corrected for
galactic extinction, {\it dereddened}, in the $r^{\prime}$, $i^{\prime}$, and
$g^{\prime}$ photometric bands.

Table~\ref{catalogA1351} lists the velocity catalogue (see also
Fig.~\ref{figottico1}): identification number of each galaxy, ID
(col.~1); right ascension and declination, $\alpha$ and $\delta$
(J2000, col.~2); $r^{\prime}$ magnitude from SDSS-DR7 (col.~3);
heliocentric radial velocities, $v=cz_{\sun}$ (col.~4) with errors,
$\Delta v$ (col.~5).

\input{catalogA1351.tex}

We had already observed A1351 field with the Wide Field Camera (WFC),
mounted at the prime focus of the 2.5 m INT telescope. We took exposures of
$9\times600$ s and $9\times300$ s in $B$ and $R$ Harris filters in
photometric conditions and 1.2\arcs seeing. However, we used SDSS-DR7
data because a greater number of photometric bands are available,
which allows an accurate colour analysis. INT and SDSS-DR7 photometric
data are very similar.  The completeness magnitude is
$r^{\prime}=20.8$.

The position of the brightest cluster member galaxy (ID.~66 with
$r^{\prime}=17.01$, hereafter BCG1) is very close to the peak of the
X-ray emission and the peak of the mass distribution provided by
\citet{dah02}.  The second-brightest galaxy is ID.~66 with
$r^{\prime}=17.73$ (hereafter BCG2), and then there are several other
galaxies at similar brightness ($\Delta r^{\prime}<0.5$).

\section{Analysis of the optical data}
\label{anal}

\subsection{Cluster member selection}
\label{memb}

\begin{figure}
\centering
\resizebox{\hsize}{!}{\includegraphics{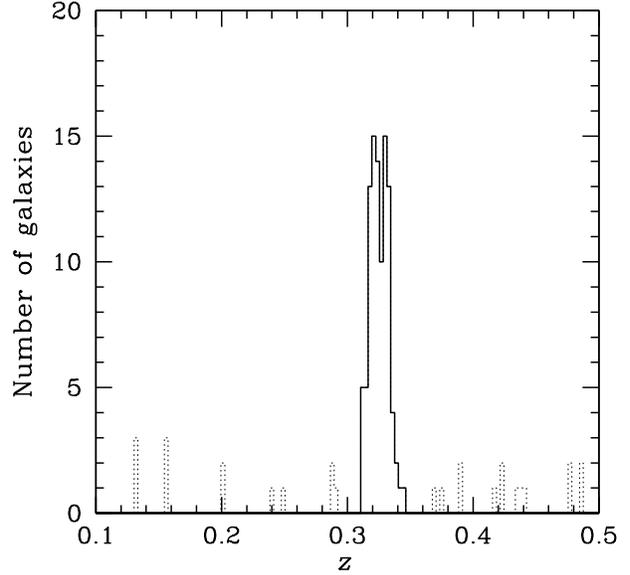}}
\caption
{Redshift galaxy distribution. The solid line histogram refers to the
100 galaxies assigned to A1351 according to the 1D-DEDICA
reconstruction method.}
\label{fighisto}
\end{figure}

To select cluster members among the 129 galaxies with redshifts we
performed the 1D adaptive-kernel method (hereafter 1D-DEDICA,
\citealt{pis93} and \citealt{pis96}; see also \citealt{fad96};
\citealt{gir96}). This procedure detected A1351 as a quite asymmetric
peak at $z\sim0.324$ populated by 100 galaxies. The re-running of the
procedure rejected another two galaxies considered as `isolated',
leading to 98 galaxies and then to the detection of two peaks (see
Figure~\ref{fighisto}). The second step in the member selection (which
uses the combination of position and velocity information, i.e.\ the
`shifting gapper' method of Fadda et al.\ (\citeyear{fad96}; see
also, e.g.\ \citealt{gir11} for details on the application of this
technique) rejects three galaxies from the whole complex. We adopted
as centre position the coordinates of the brightest cluster galaxy --
BCG1 [R.A. = $11^{\mathrm{h}}42^{\mathrm{m}}24\dotsec78$,
  Dec. = $+58\degree 32\arcmm 05.3\arcs$ (J2000.0)], which coincides
with the location of the peak of the X-ray emission. However, note
that A1351 is an elongated cluster with BCG1 lying in the northern
region (see in the following). Summarizing, we analysed the structure
of A1351 using 95 fiducial cluster members.

\begin{figure}
\centering
\resizebox{\hsize}{!}{\includegraphics{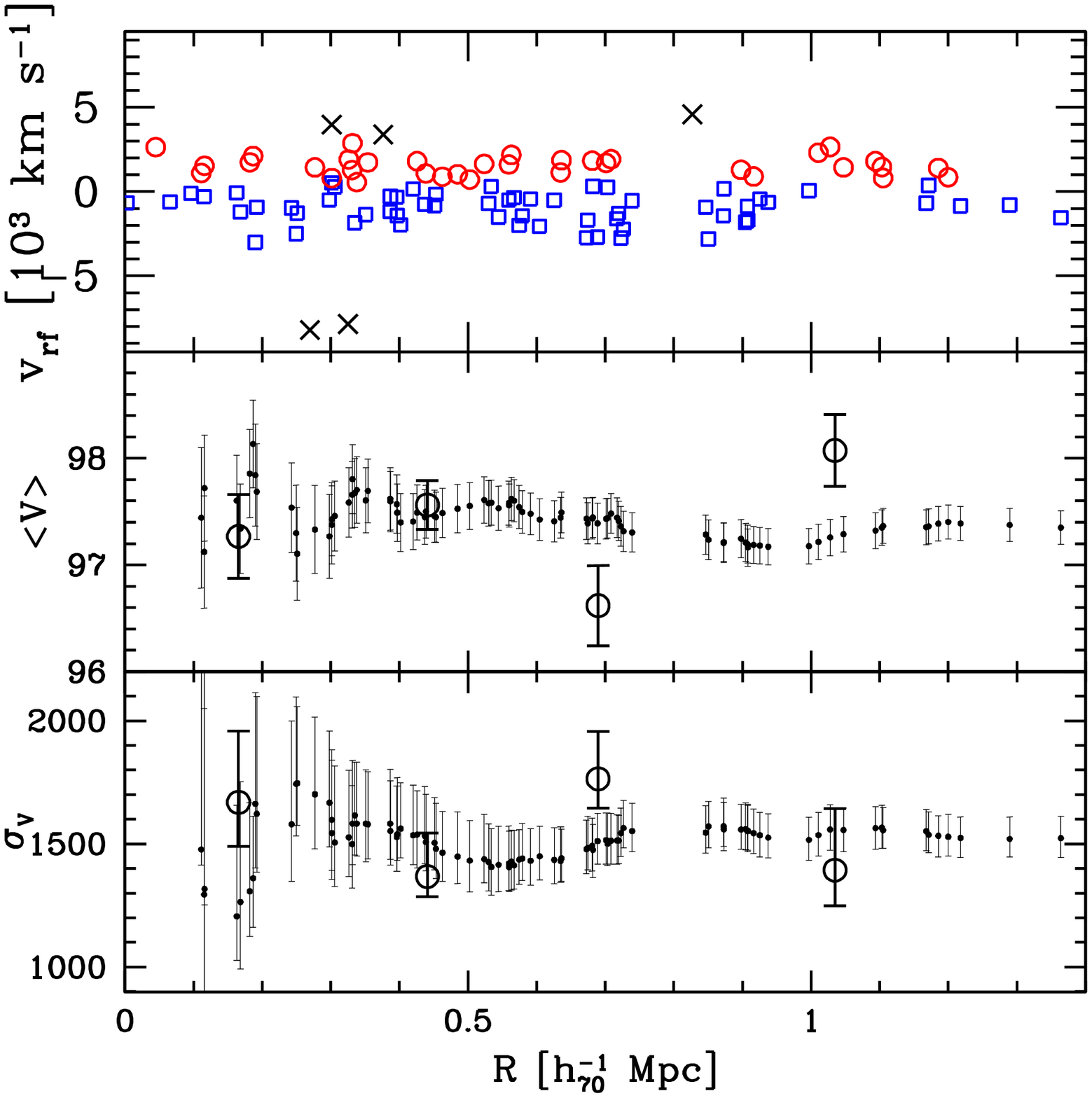}}
\caption
{{\em Top panel:} rest-frame velocity vs.\ projected clustercentric
  distance for the 100 member galaxies belonging to the peak in the
  velocity distribution (Fig.~\ref{fighisto}). Crosses indicate
  galaxies rejected as interlopers by the `shifting gapper'
  procedure. Blue squares and red circles indicate the 95 member
  galaxies as divided in the DED1 and DED2 subsamples. The cluster
  centre coincides with the position of the BCG1. {\em Middle and
    bottom panels:} differential (big circles) and integral (small
  points) profiles of mean velocity and LOS velocity dispersion,
  respectively. For the differential profiles, we plot the values for
  four annuli from the centre of the cluster, each of 0.3 \hh. For the
  integral profiles, the mean and dispersion at a given (projected)
  radius from the cluster centre is estimated by considering all
  galaxies within that radius -- the first value computed on the five
  galaxies closest to the centre. The error bands at the $68\%$
  c.l.\ are also shown.}
\label{figprof}
\end{figure}

By applying the biweight estimator to the 95 cluster members
(\citealt{bee90}, ROSTAT software), we computed a mean cluster
redshift of $\left<z\right>=0.3247\pm$ 0.0005, i.e.\ 
$\left<v\right>=(97\,350\pm$ 157) \kss. We estimated the LOS velocity
dispersion, $\sigma_{\rm V}$, by using the biweight estimator and
applying the cosmological and standard corrections for velocity errors
\citep{dan80}, respectively. We obtained $\sigma_{\rm
  V}=1524_{-74}^{+96}$ \kss, where the errors are computed through a
bootstrap technique. To verify the robustness of the $\sigma_{\rm V}$
estimate, we analysed the velocity dispersion profile, which is
consistent with a flat shape (Fig.~\ref{figprof}).

\subsection{Substructure analysis}
\label{sub}

\begin{figure}
\centering 
\resizebox{\hsize}{!}{\includegraphics{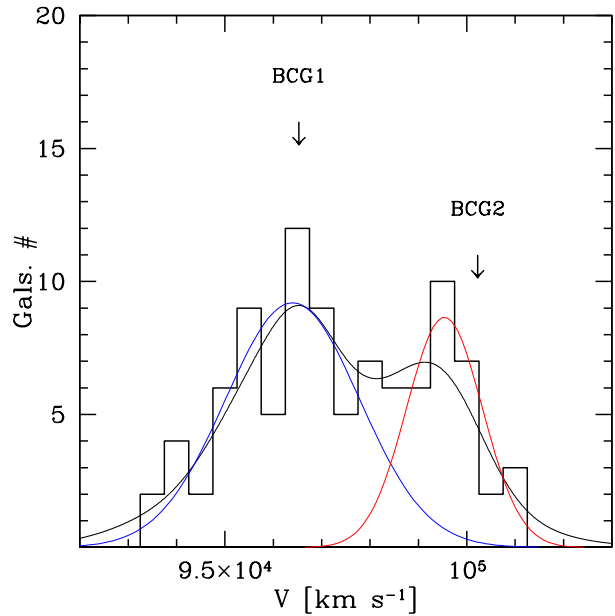}}
\caption
{Velocity distribution of the 95 galaxies assigned to the cluster.
  The arrows indicate the velocity of the BCG1 and BCG2. Black curve
  is the density reconstruction according to the 1D-DEDICA method. Red
  and blue Gaussian profiles are the best bimodal fits according to
  the 1D-KMM test.}
\label{figgauss}
\end{figure}

In the velocity distribution of the 95 member galaxies, the 1D-DEDICA
procedure detects two significant peaks (DED1 and DED2) of 60 and 35
galaxies, at $96\,525$ and $99\,134$ \kss. These peaks largely overlap
with 22 (of DED1) and 19 (of DED2) galaxies with a high probability
of belonging to the other peak too, see Fig.~\ref{figgauss}. BCG1 and
BCG2 belong to the DED1 and DED2 subclusters, respectively. Mean
velocities and velocity dispersions, as computed on subcluster
members, are $\left<v_{\rm DED1}\right>=96\,191$ \ks and $\left<v_{\rm
  DED2}\right>=99\,433$ \kss and $\sigma_{v,\rm DED1}=940$ \ks and
$\sigma_{v,\rm DED2}=584$ \kss.  The distributions of galaxy positions
of DED1 and DED2 are different at the 99.4\% c.l.\ according to the the
2D Kolmogorov--Smirnov test (\citealt{fas87}, hereafter 2D-KS), see
Figure~\ref{figds10v}.

The 1D-Kaye mixture model test (1D-KMM; \citealt{ash94}; see also
\citealt{bos12}, \citealt{bos13} for recent applications), finds a
two-Gaussian partition which is a better descriptor than a one-Gaussian
profile at the 98.4\% and 96.5\% c.l.s for the homoscedastic and
heteroscedastic cases (i.e. the cases assuming equal or
different dispersions), respectively. The Gaussians corresponding
to the heteroscedastic case, which is the more realistic approach, are
shown in Figure~\ref{figgauss}. The two subsamples detected by 1D-KMM
contain 62 and 33 members (KMM1 and KMM2), where several galaxies have
a high probability of belonging to both the peaks. For the two Gaussians
we obtain mean velocities, $\left<v_{\rm KMM1}\right>=96\,328$ \ks and
$\left<v_{\rm KMM2}\right>=99\,538$ \kss, and velocity dispersions,
$\sigma_{v,\rm KMM1}=1\,384$ \ks and $\sigma_{v,\rm KMM2}=783$ \kss,
where the galaxies are weighted according to their partial membership
of both groups so as to avoid possible underestimates connected with
an artificial truncation of the tails of the samples. The
distributions of the galaxy positions of KMM1 and KMM2 are different at
the 99.3\% c.l.\ according to the the 2D-KS.
  
\begin{figure}
\centering
\resizebox{\hsize}{!}{\includegraphics{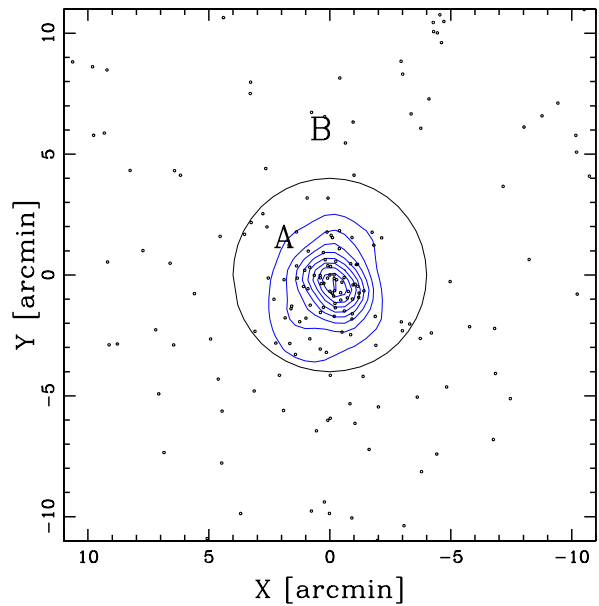}}
\caption
{A large scale spatial distribution on the sky and relative isodensity
  contour map of likely cluster members extracted from the SDSS
  photometric catalogue with $r^{\prime}\le 21.5$. The contour map is
  obtained with the 2D-DEDICA method (blue lines). The $4^{\prime}$
  circle outlines the region sampled by our spectroscopic data. The
  two labels indicate the two secondary peaks in the mass distribution
  according to the weak lensing analysis by \citet{hol09}. }
\label{figk2}
\end{figure}

\begin{figure}
\centering 
\resizebox{\hsize}{!}{\includegraphics{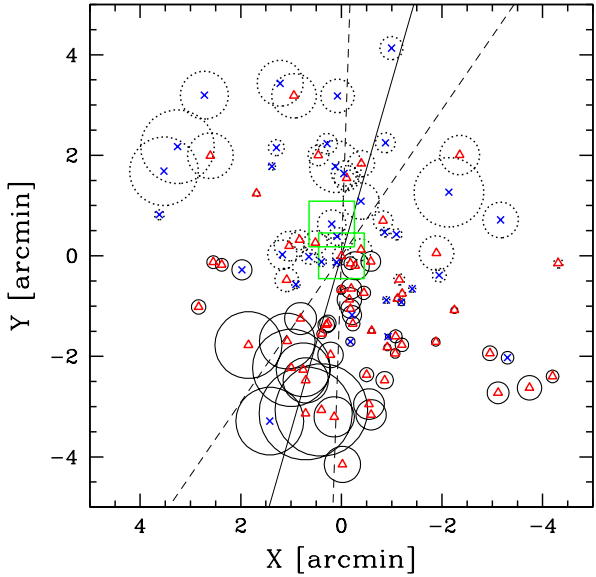}}
\caption
{Spatial distribution of the 95 cluster members, where small blue
  triangles and red crosses indicate the DED1 and DED2 members.  Each
  galaxy is marked by a circle: the larger the circle, the larger is
  the deviation of the local mean velocity from the global mean
  velocity.  Thin/solid and thick/dotted circles show where the local
  value is smaller or larger than the global value. The plot is
  centred on the cluster centre (=BCG1). BCG1 and BCG2 are indicated
  by faint large/green squares. The solid and dashed lines indicate
  the position angle of the cluster velocity gradient and relative
  errors, respectively.}
\label{figds10v}
\end{figure}

Concerning the spatial distribution of galaxies in the plane of the sky,
we analysed the photometric SDSS catalogues extracted for a 30\arcmin
radius region from the cluster centre thus to overcome redshift and
spatial incompleteness of our spectroscopic catalogue. We performed a
member selection based on both ($r^{\prime}$--$i^{\prime}$
vs.\ $r^{\prime}$) and ($g^{\prime}$--$r^{\prime}$ vs.\ $r^{\prime}$)
colour--magnitude relations following, for example, the \citet{bos12} prescriptions;
here $r^{\prime}$--$i^{\prime}$=0.956-0.019$r^{\prime}$ and
$g^{\prime}$--$r^{\prime}$=2.841-0.244$r^{\prime}$. The 2D-DEDICA
analysis detects A1351 as only one isolated peak in the space of
projected positions (see Figure~\ref{figk2}).

We also applied a set of 3D tests combining velocity and position
information.  We applied the $\Delta$-statistics devised by Dressler
\& Schectman (\citeyear{dre88}, hereafter DS-test), also with the
alternative kinematic estimators separating mean velocity and
velocity dispersion (hereafter DS$\left<v\right>$-test and
DS$\sigma_{v}$-test, e.g.\ \citealp{fer03,gir10,gir97}). We
applied the $\epsilon$-test \citep{bir93} based on the projected mass
estimator and the centroid shift or $\alpha$-test \citep{wes90}. The
presence of a velocity gradient was analysed performing a multiple
linear regression fit to the observed velocities with respect to the
galaxy positions in the plane of the sky (e.g.\
\citealt{den96}). The significance of the results is based on 1000
Monte Carlo simulated clusters where original position and velocities
are shuffled. We found that substructure is significant in the case
of the the $\epsilon$-test, which is sensitive to the combined values
of local velocity dispersion and projected density at the $>99.9\%$
c.l. The DS-test detects substructure significant at the $98.8\%$
c.l., in practice owing to the mean velocity estimator (for DS$\left< v
\right>$-test the significance is at the 99.7\% c.l.). We also found a
$>99.9\%$ c.l.\ significant velocity gradient with a position angle 
$PA = -16_{-18}^{+14}$ degrees (measured counter-clockwise from the
north), i.e.\ higher-velocity galaxies lie in the north (see
Figure~\ref{figds10v}). Figure~\ref{figds10v} shows that galaxies in the
external cluster region are responsible for the position--velocity
correlation.

\begin{figure}
\centering 
\resizebox{\hsize}{!}{\includegraphics{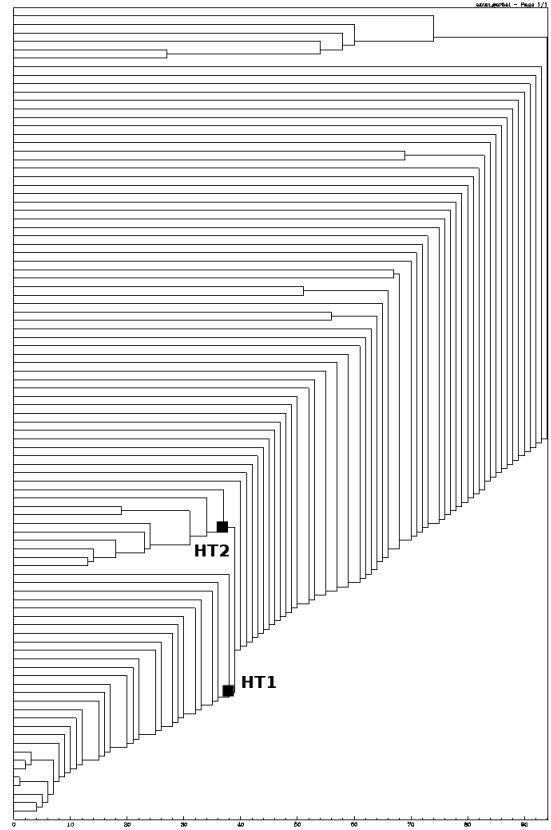}}
\caption
{Dendogram obtained through the \citet{ser96} algorithm (here
$M/L_r$=150 \mll). The abscissa is the binding energy (here in
arbitrary unit with the deepest negative energy levels on the left).
The two main subclusters are indicated by HT1 and HT2.}
\label{a1351gerbal}
\end{figure}

\begin{figure}
\centering
\resizebox{\hsize}{!}{\includegraphics{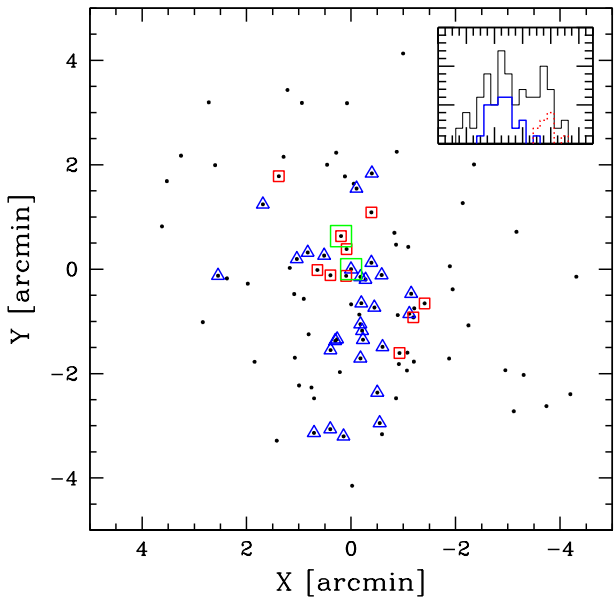}}
\caption
{Spatial distribution on the sky of 95 cluster members (black points).
  Blue triangles and red squares indicate galaxies of HT1, the main
  subcluster, and HT2, the secondary subcluster. BCG1 and BCG2 are
  highlighted by large, green squares. The insert plot shows the same
  velocity distribution of subcluster galaxies (blue/solid dotted and
  red/dotted lines for HT2 and HT1).}
\label{fight}
\end{figure}

We also resorted to the Htree-method devised by \citet{ser96}; a
similar technique is also used to determine cluster members
\citep{ser13}. The method uses a hierarchical clustering analysis to
determine the relationship between galaxies according to their
relative binding energies where ignorance of the mass
associated with each galaxy is overcome adopting for each galaxy halo
the typical mass-to-light of galaxy cluster. The clustering
sequence can be visualized in the form of a hierarchical tree
diagram (dendogram). Figure~\ref{a1351gerbal} shows the resulting
dendogram, where we used $M/L_r$=150 \mll, as suggested by broad
statistical studies (e.g.\ \citealt{gir00,pop05}). From
Fig.~\ref{a1351gerbal,} it can be seen that A1351 contains two main
subgroups (HT1 and HT2), with BCG1 and BCG2 in the respective
potential well.  The presence of the two subgroups is quite
robust against the choice of the value of $M/L_r$, although the
number of identified members depends on the precise value of
$M/L_r$, i.e.\ it decreases with the increase in $M/L_r$ from 100 to
200 \mll. HT1 and HT2 are well separated in velocity and
superimposed in the sky in the central cluster region (see
Fig.~\ref{fight} and insert plot). Their mean velocities are
$\left<v_{\rm HT1}\right>=96\,540$ \ks and $\left<v_{\rm
  KMM2}\right>=99\,697$ \kss.  Our interpretation is that they are
the central, densest parts of the two subclusters detected through
the KMM or the 1D-DEDICA method (see above).

\section{X-ray morphological and spectral analysis}
\label{Xmorph}

In order to complement our optical analysis, we used recent X-ray data
taken with {\it Chandra} and {\it XMM-Newton} satellites to study the
ICM of A1351 and infer its global morphological and spectral
properties.

First, we considered archival data taken with the {\it Chandra}
Advanced CCD Imaging Spectrometer (ACIS--I; exposure ID~15136). We
reduced the data using the {\small CIAO} package\footnote{see
  http://asc.harvard.edu/ciao/} (version 4.2) on chips I0, I1, I2 and
I3 (total field of view $\sim 17\arcmin\times 17\arcmin$) in a standard way
(see \citealt{bos04})

\begin{figure}
\centering
\resizebox{\hsize}{!}{\includegraphics{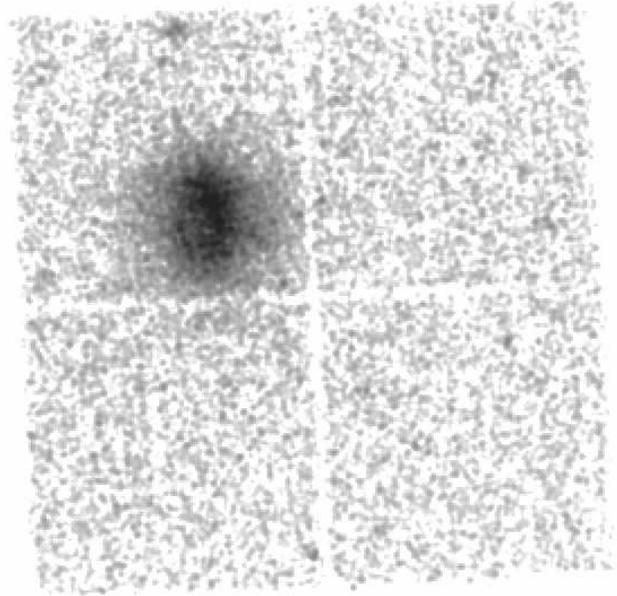}}
\caption
{17\arcmm$\times$17\arcm {\it Chandra} ACIS-I X-ray image (ID~15136)
of A1351 in the energy band 0.3--7 keV (north at the top and east to
the left; binned by 8$\times$8 pixels and Gaussian-smoothed with 
$\sigma$=2 pixels).}
\label{figX1}
\end{figure}

A quick look at the reduced image suggests the ICM is somewhat
elongated in the NNE--SSW direction (see Figure~\ref{figX1}). A clear
twisting of the isophotes is present (see Fig.~\ref{figimage1}), from
the centre towards the more external regions, thus suggesting the ICM
is not relaxed in a regular gravitational potential well.

\begin{figure}
\centering
\resizebox{\hsize}{!}{\includegraphics{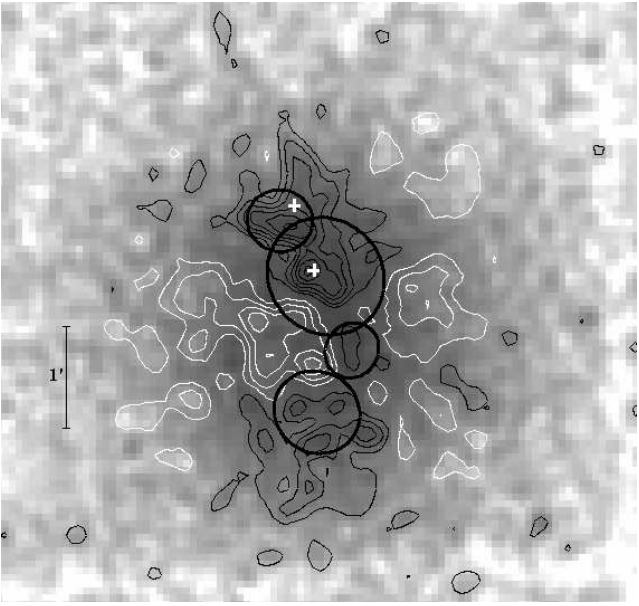}}
\caption
{The smoothed X-ray emission in the left-upper quadrant of
Fig.~\ref{figX1} with, superimposed, the contour levels of the
positive (black thin contours) and negative (white thin contours)
smoothed $\beta$-model residuals (north at the top and east to the
left). Thick ellipses represent the X-ray clumps found by the
wavelet multiscale analysis. As a reference, crosses highlight the
positions of BCG1 and BCG2.}
\label{residui}
\end{figure}

To better characterize the X-ray morphology of the cluster we
performed a wavelet multiscale analysis on chip I3 with the 
{\small CIAO/WAVDETECT} task. On small scales (2--6 pixels), taking
advantage of the high spatial resolution of {\it Chandra} images,
{\small WAVDETECT} was effective in identifying the pointlike sources
projected on the line of sight of the cluster. After removal of
these sources, a run of {\small WAVDETECT} on larger scales (20--42 
pixels) allowed us to identify four X-ray clumps (see
Fig.~\ref{residui}), thus confirming the irregular morphology of the
ICM.

These clumps are also evident when fitting a simple $\beta$-model
profile to the 2D X-ray photon distribution and analysing the
model residuals. The model is defined to be (Cavaliere \&
Fusco-Femiano \citeyear{cav76}):
\begin{equation}
S({\rm R})=S_0[1+({\rm R}/{\rm R}_{\rm c})^2]^{\alpha}+b,
\end{equation}
where $\rm R$ is the projected radial coordinate from the centroid
position and $b$ the surface brightness background level. The fit only
provides poorly constrained and unrealistic values of the slope
parameter $\alpha$ and the angular core radius R$_{\rm c}$ (
  $\alpha=-4.4^{+0.5}_{-0.4}$, R$_{\rm
    c}=163.1\arcs^{+12.6\arcs}_{-10.7\arcs}=766.6^{+59.2}_{-50.2}$
  \kpc at the cluster redshift), which is another hint of an
irregular X-ray photon distribution. Moreover, the analysis of the
$\beta$-model residuals reveals a clear excess of X-ray emitting gas
in correspondence with the four clumps detected by the wavelet
multiscale analysis (see Figure~\ref{residui}).

As for the spectral properties of the cluster X-ray photon
distribution, we computed global ICM temperature by analysing {\it
  XMM-Newton} data. We considered an observation of A1351
(ID~0650382201) stored in the public archive. We reprocessed the
Observation Data Files (ODF) using the Science Analysis System (SAS;
version 13.0), following the procedure described in \citet{deg09} to
produce calibrated event list cleaned of soft protons. The net
exposure time of the observation is 5.2 ks for MOS1, 6.7 ks for MOS2
and 2.5 ks for the {\it pn}.

\begin{figure}
\centering
\resizebox{\hsize}{!}{\includegraphics{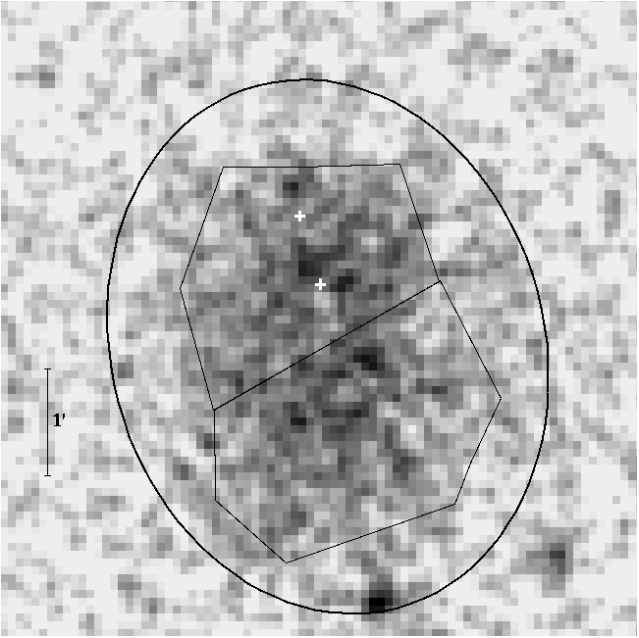}}
\caption
{{\it XMM} smoothed image (ID~0650382201) of A1351 in the energy band
2--10 keV (north at the top and east to the left). Ellipse and
polygons define the regions considered for the measurement of the ICM 
temperatures. As a reference, crosses show the positions of BCG1 and 
BCG2.}
\label{figX2}
\end{figure}

We applied the background subtraction using blank-sky fields for EPIC
MOS and {\it pn} produced by \citet{lec08}. Using the {\small XSPEC}
package (version 11.3.2, \citealt{arn96}), we analysed the EPIC
spectrum of an elliptical region centred on coordinates
$\alpha$ = 11:42:24.28 and $\delta$ = +58:31:30.94, with $a=2.5^\prime$,
$a/b$=1.25 (where $a$ and $b$ are the major and minor axes,
respectively) and $PA$=15\degree (measured from north to east; see
Figure~\ref{figX2}). We modelled the spectrum with a 
one-temperature
thermal model with the plasma in collisional ionization equilibrium
({\it apec} model in {\small XSPEC}), multiplied by a {\it wabs}
absorption model with the Galactic hydrogen column density fixed at
$N_{\rm H}=9.94\times10^{19}$ cm$^{-2}$ \citep{kal05}. We found that
the ICM has a global temperature of $kT=8.62^{+0.96}_{-0.61}$ keV.

\section{Cluster structure}
\label{disc}

We present several proofs in favour of the previous first evidence of
an ongoing merger activity in A1351. Our kinematic analysis is based 
on a factor $>5$ more galaxies than previously. On the one hand, we confirm 
that A1351 is characterized by a high global value of 
velocity dispersion $\sigma_{\rm V}=1524_{-74}^{+96}$ \ks (see 
$\sigma_v=1680_{-230}^{+340}$ \kss, \citealt{irg02}) and, on the 
other hand, it shows a bimodal velocity distribution with two
peaks separated by $\sim$2500 \ks in the rest frame. As for the X-ray
data, we confirm the N--S elongation of the X-ray emission and further
features are detected. From north to south we detected three, maybe
four, substructures (see Figure~\ref{residui}). The northern part,
containing the two most important X-ray clumps, is particularly
brighter and is NNE--SSW elongated. BCG2 and BCG1 bright galaxies
are placed along the N--S direction. In addition, a few other quite
luminous members (IDs.~55, 56, 76, 79) are aligned following the same 
axis. Moreover, we detected a velocity gradient in the galaxy
velocity field roughly in S--N direction and the presence of 3D
substructure at a very significant level.

The above results indicate that the cluster structure extends along
the LOS but is not completely aligned with it. In particular, the
two main subclusters, corresponding to BCG1 and BCG2 and detected as
two northern X-ray clumps, cause the bimodality of the velocity
distribution, low and high velocity peaks respectively (see 
Fig.~\ref{figgauss}). In spite of the very significant 3D substructure 
tests, we cannot detect two separate galaxy concentrations. This is 
probably because the two subclusters are so closely projected on the 
sky that we would need a much larger number of redshifts to 
sample the region better. The southern region, mostly populated by galaxies 
of the low velocity peak, is instead characterized by two less 
important substructures that contribute to trace the north--south direction.

The high value of the X-ray temperature, $kT=8.69^{+1.01}_{-0.54}$
keV, indicates a high mass value, e.g.\  $M_{500}\sim 1.1$ \mqui and
$M_{200}\sim 1.8$ \mqui \citep{lop09}, as obtained from two published
$M$--$T$ relations. Alternatively, we used a kinematic approach, taking
advantage of our redshift sample. We followed the prescriptions of
Girardi \& Mezzetti (\citeyear{gir01}, see also \citealt{gir98}) by
considering that A1351 is probably formed by two subclusters: the low
velocity one, with $\sigma_{v,1}=900$-$1400$ \kss, and the
high-velocity one, with $\sigma_{v,2}=600$-$800$ \kss.  We obtained
$M_1(<R_{\rm 200,1}=1.9-2.8 \hhh)=1.1-3.6$ \mqui and $M_2(<R_{\rm
  200,2}=1.2-1.6 \hhh)=0.3-0.6$ \mquii, and a global value of
$M_{\rm sys}=1.4$-4.2 \mquii. A1351 is thus quite a massive system, in
agreement with typical mass values of other clusters hosting radio
haloes (e.g.\ \citealt{gir08}; \citealt{gir10}) and in line with the
expectations of the re-acceleration scenario in favouring more massive
systems (e.g.\ \citealt{cas08}).

The value of the velocity dispersion of the main subcluster,
$\sigma_{v,1}=900$-$1400$ \kss, is consistent with the $\sigma_{\rm
  v,SIS}$ value of the singular isothermal sphere in the weak
gravitational analysis ($\sigma_{v,1}=1000$-$1400$ \kss, Dahle et
al.\ \citeyear{dah02}; Holjem et al.\ \citeyear{hol09}). The lower limit
of our mass estimate, $M_{\rm sys}=1.4$--4.2 \mquii, is only slightly
above the lensing mass of Holjem et al.\ (\citeyear{hol09}), $M_{\rm
  200}=8$-9 \mquaa, when taking into account their smaller estimate of the value of
$R_{200}$ (1.69 \hh, see our above estimate of $R_{\rm 200,1}$).
Instead, we disagree with the morphological details of the two
previous gravitational lensing analyses of \citet{dah02} and
\citet{hol09}, which, in fact, are in disagreement with each other (see
Section~\ref{intro}).  In particular, as for the northern minor peak
`B' detected by \citet{hol09}, this is outside the field we sampled
through TNG redshift data and it is not detected by our analysis of
the SDSS photometric sample (see Fig.~\ref{figk2}), where the presence
of A1351 is outstanding. Indeed, \citet{hol09} have suggested the
possibility that peak `B' is only a chance alignment of
galaxies. The important LOS elongation of the structure of A1351 might be
the cause the difficulties in the weak gravitational lensing analysis,
in particular in determining the concentration parameter
\citep{hol09}.

The elongation and twisting of isophotes of the X-ray emission is
strong evidence for an ongoing cluster merger. With the available {\it
  XMM-Newton} observation we did not find any X-ray temperature
difference between northern and southern regions (see
Figure~\ref{figX2}).

\begin{figure}
\centering
\resizebox{\hsize}{!}{\includegraphics{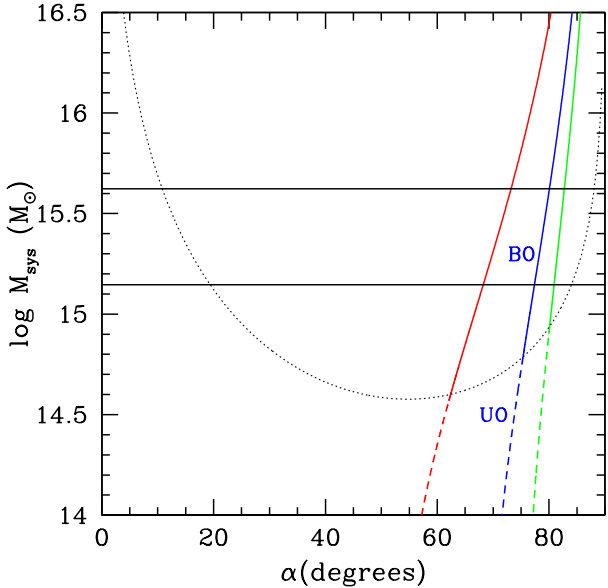}}
\caption
{System mass vs.\ projection angle diagram of the two-body model
applied to the two (main) subclusters. Solid and dashed lines correspond
to bound {bf (BO) and unbound  (UO) outgoing solutions, respectively,
where the elapsed time from the core crossing is
$t=0.1$, 0.2 and 0.3  Gy (from left to right, i.e.\ red, blue and green
lines).} The total mass
of the system is represented by the horizontal line. The Newtonian
criterion separates bound and unbound regions, above and below the 
thin dotted curve, respectively.}
\label{figbim}
\end{figure}

Reducing the complex structure of A1351 to the two main subclusters,
we can apply a model to follow the merging kinematics. We applied the
analytic method of the bimodal model (\citealt{bee82};
\citealt{tho82}; see also \citealt{gir08} for a recent
application). As for the input values of the relevant parameters for
this model, we used the above $M_{\rm sys}=1.4$-4.2 \mqui and the
rest-frame LOS velocity difference between the two subclusters that
measured $\Delta V_{\rm rf,LOS}\sim$
2500 \kss in both the DEDICA and KMM analyses . As for the projected distance between the two subclusters,
our substructure analysis suggests a small value and does not allow a
precise estimate. We then resorted to that measured between BCG1 and
BCG2, $D\sim 0.2$ \hh. In other clusters hosting radio haloes the time,
$t$, elapsed from the core crossing is $t=0.1$--0.3 Gyr (e.g.\ 
\citealt{mar02}; \citealt{gir08}). For the cases with $t=0.1$,
  0.2 and 0.3 Gy, Figure~\ref{figbim} shows the outgoing
  solutions of the bimodal model as a function of $\alpha$, the
projection angle between the plane of the sky and the line connecting
the centres of the two clumps, with the observed mass of the system
$M_{\rm sys}$.  The post core-crossing scenario represented by the
bound outgoing solutions (BO) leads to a range of $\alpha
\magcir 65$--80\degree. This value agrees with observational
evidence of two main subclusters aligned mostly, but not completely, along
the LOS.

Regarding the diffuse radio emission and its peculiar asymmetry with
respect to the X-ray peak, \citet{gia09} have suggested that the
southern X-ray emission, the `ridge', might be a relic rather a
southern extension of the radio halo. Our results strongly support the
above hypothesis. In fact, the northern part of the radio emission
coincides with the region where the two main subclusters are
projected.  Consequently, the northern part of the radio emission can
be correctly interpreted as a radio halo centrally located and someway
symmetric with respect to the X-ray central emission, in agreement
with typical radio halos in clusters. Excluding the contribution of
the ridge, A1351 is much closer to the relation between the radio
power and X-ray luminosity, as traced by the clusters showing a giant
radio halo, as is shown well in fig.~3 of \citet{gia09}. The
southern part of the radio emission (the ridge) lies at the external
border of the most southern X-ray clumps. Thus, it could most probably be
a relic connected to a merger along the north--south direction
involving minor galaxy clumps.  The only (minor) point of their
scenario we cannot agree with is the northern small radio filament (see
their figure~2). In this respect, we found no extension in the X-ray
emission and no evidence of an optical subcluster in that region.

\section{Summary and conclusions}
\label{concl}

Our analysis of A1351 at $z\sim 0.3$, based on new spectra
acquired at the TNG for $\sim$100 member galaxies, and X-ray data
retrieved from {\it Chandra} and {\it XMM} archives, derives the
following results:

\begin{itemize}
\item the measured dispersion of galaxy velocities is so high,
$\sigma_{\rm V}\sim 1500$ \kss, owing to the bimodal structure
of the velocity distribution;

\item the measured X-ray temperature, $kT\sim 9$ keV, is also high,
and both X-ray and kinematic data lead to a high mass value 
(1--4 \mquii), in agreement with observational estimates of most 
clusters hosting radio haloes and with theoretical expectations 
of the reacceleration/turbulence scenario;

\item A1351 is a very complex cluster, with two main subclusters
projected in the northern region and connected with the two peaks in
the velocity distribution, and other minor substructures, all
tracing the N--S cluster elongation.
\end{itemize}

Therefore, we conclude that in the above scenario, the quite peculiar,
diffuse radio emission can be reduced to a much more typical emission
(halo plus relic), strongly supporting a previous suggestion 
based only on the inspection of radio data and preliminary X-ray data.  We
stress, however, that the study of so complex a cluster would be greatly advantageous with a substantial improvement of data. A much larger
redshift dataset would allow the detection of individual subcluster
cores. In addition, deeper X-ray observations would make it possible to
construct X-ray temperature and metallicity maps.

More generally, considering our previous results about structure of
DARC clusters, we conclude that peculiar features in the diffuse radio
emission could be interpreted by considering the details of the complex
cluster structure, for instance in Abell 520, hosting a very elongated
radio halo \citep{gir08}, and in Abell 2345, hosting two quite
asymmetric radio relics \citep{bos10}. In this context, the study of
the galaxy kinematics represents an important ingredient, critical to our 
understand of the geometry of the merger and to detecting possible LOS
aligned substructures.

\section*{Acknowledgements}

We are indebted to Tiziana Venturi for the VLA radio image.  RB
acknowledges financial support from the Spanish Ministry of Economy
and Competitiveness (MINECO) under the 2011 Severo Ochoa Programme
MINECO SEV-2011-0187. MG acknowledges financial support from MIUR
PRIN2010-2011 (J91J12000450001). This publication is based on
observations made on the island of La Palma with the Italian
Telescopio Nazionale Galileo (TNG) and the Isaac Newton Telescope
(INT).  The TNG is operated by the Fundaci\'on Galileo Galilei -- INAF
(Istituto Nazionale di Astrofisica). The INT is operated by the Isaac
Newton Group. Both telescopes are located in the Spanish Observatorio
del Roque de Los Muchachos of the Instituto de Astrofisica de
Canarias.  This research has made use of the NASA/IPAC Extragalactic
Database (NED), which is operated by the Jet Propulsion Laboratory,
California Institute of Technology, under contract with the National
Aeronautics and Space Administration.

\end{document}

%% file: catalogA1351.tex

\begin{table}
        \caption[]{Radial Velocities of 135 galaxies in the field of
          A1351. IDs in boldface highlight the two brightest cluster
          members BCG1 (ID. 66) and BCG2 (ID 74; see text). IDs in
          italics refer to nonmember galaxies.}
         \label{catalogA1351}
              $$ 
           \begin{array}{r c r r r}
            \hline
            \noalign{\smallskip}
            \hline
            \noalign{\smallskip}

\mathrm{ID} & \alpha,\delta\,(\mathrm{J}2000) & r^{\prime}\,\,\, & v\,\,\,\,\,& \Delta v\\
 &                  & &\mathrm{\,(\,km}&\mathrm{s^{-1}\,)}\\
            \hline
            \noalign{\smallskip}  

  1 &  11\ 41\ 51.74 ,+58\ 31\ 56.4 &   20.36 &  96298 &  86  \\  
  2 &  11\ 41\ 52.65 ,+58\ 29\ 41.4 &   19.10 &  95376 &  77  \\  
  3 &  11\ 41\ 56.16 ,+58\ 29\ 27.7 &   19.78 &  96378 &  53  \\  
\textit{4} &  11\ 41\ 58.06 ,+58\ 32\ 11.0 &   20.11 & 131414 & 101  \\  
\textit{5} &  11\ 41\ 59.35 ,+58\ 31\ 33.0 &   20.61 & 117212 &  79  \\  
  6 &  11\ 41\ 59.49 ,+58\ 30\ 03.6 &   19.85 &  99815 &  90  \\  
  7 &  11\ 42\ 00.51 ,+58\ 32\ 48.1 &   20.14 &  98618 & 123  \\  
  8 &  11\ 42\ 00.94 ,+58\ 29\ 21.8 &   19.69 &  96522 &  66  \\  
\textit{9} &  11\ 42\ 01.74 ,+58\ 28\ 57.0 &   21.58 & 223024 & 220  \\  
 10 &  11\ 42\ 02.17 ,+58\ 30\ 09.0 &   19.26 &  97493 &  48  \\  
\textit{11} &  11\ 42\ 05.93 ,+58\ 30\ 40.2 &   20.10 &  18684 & 158  \\  
 12 &  11\ 42\ 06.73 ,+58\ 34\ 05.6 &   21.03 &  97644 & 119  \\  
 13 &  11\ 42\ 07.57 ,+58\ 31\ 00.6 &   20.76 &  97757 &  59  \\  
\textit{14} &  11\ 42\ 07.93 ,+58\ 29\ 36.0 &   21.79 & 404991 & 500  \\  
\textit{15} &  11\ 42\ 08.05 ,+58\ 27\ 19.5 &   21.19 & 126950 &  62  \\  
\textit{16} &  11\ 42\ 08.08 ,+58\ 32\ 25.4 &   20.07 &  60602 &  81  \\  
 17 &  11\ 42\ 08.39 ,+58\ 33\ 21.2 &   18.66 &  99683 &  75  \\  
 18 &  11\ 42\ 09.88 ,+58\ 31\ 42.1 &   21.18 &  99562 &  95  \\  
 19 &  11\ 42\ 10.29 ,+58\ 32\ 08.6 &   20.77 &  97823 &  79  \\  
 20 &  11\ 42\ 10.41 ,+58\ 30\ 22.5 &   18.89 &  95304 &  39  \\  
\textit{21} &  11\ 42\ 13.67 ,+58\ 31\ 50.9 &   20.14 & 116817 & 114  \\  
 22 &  11\ 42\ 13.99 ,+58\ 31\ 25.9 &   18.59 &  98855 &  37  \\  
 23 &  11\ 42\ 15.51 ,+58\ 31\ 20.3 &   19.82 &  94808 &  70  \\  
 24 &  11\ 42\ 15.57 ,+58\ 30\ 19.0 &   20.65 &  94702 &  88  \\  
 25 &  11\ 42\ 15.64 ,+58\ 31\ 09.9 &   18.85 &  99819 &  84  \\  
 26 &  11\ 42\ 15.96 ,+58\ 31\ 37.1 &   19.88 &  95611 &  70  \\  
 27 &  11\ 42\ 16.24 ,+58\ 31\ 14.1 &   20.41 &  96996 &  81  \\  
\textit{28} &  11\ 42\ 16.28 ,+58\ 35\ 37.0 &   21.59 &  19432 & 156  \\  
 29 &  11\ 42\ 16.39 ,+58\ 32\ 30.9 &   19.79 & 101230 &  70  \\  
 30 &  11\ 42\ 16.52 ,+58\ 30\ 29.3 &   21.43 &  95433 & 110  \\  
 31 &  11\ 42\ 16.59 ,+58\ 30\ 08.8 &   20.70 &  96742 &  86  \\  
\textit{32} &  11\ 42\ 16.77 ,+58\ 31\ 42.3 &   19.99 &  86266 &  68  \\  
 33 &  11\ 42\ 17.13 ,+58\ 36\ 13.2 &   19.83 &  98554 &  66  \\  
 34 &  11\ 42\ 17.68 ,+58\ 30\ 28.9 &   19.44 &  99607 &  57  \\  
 35 &  11\ 42\ 17.77 ,+58\ 30\ 16.2 &   20.33 &  94769 &  81  \\  
 36 &  11\ 42\ 17.93 ,+58\ 31\ 12.5 &   20.61 &  99725 & 114  \\  
 37 &  11\ 42\ 18.07 ,+58\ 34\ 20.2 &   20.66 &  99856 &  77  \\  
 38 &  11\ 42\ 18.18 ,+58\ 32\ 33.5 &   20.63 &  99328 &  73  \\  
 39 &  11\ 42\ 18.19 ,+58\ 29\ 36.9 &   21.14 &  96715 & 101  \\  
 40 &  11\ 42\ 18.42 ,+58\ 32\ 47.1 &   20.47 &  97793 & 147  \\  
\textit{41} &  11\ 42\ 19.73 ,+58\ 29\ 15.6 &   19.73 &  71610 & 130  \\  
\textit{42} &  11\ 42\ 19.77 ,+58\ 29\ 32.2 &   20.31 & 126880 & 110  \\  
 43 &  11\ 42\ 20.17 ,+58\ 30\ 36.0 &   20.08 &  97199 &  70  \\  
          
            \noalign{\smallskip}			 
            \hline					    
            \noalign{\smallskip}			    
            \hline					    
         \end{array}					 
     $$ 						 
         \end{table}					 
\addtocounter{table}{-1}				 
\begin{table}					 
          \caption[ ]{Continued.}
     $$ 
           \begin{array}{r c r r r}
            \hline
            \noalign{\smallskip}
            \hline
            \noalign{\smallskip}

\mathrm{ID} & \alpha,\delta\,(\mathrm{J}2000) & r^{\prime}\,\,\, & v\,\,\,\,\,& \Delta v\\
 &                 & &\mathrm{(\,km}&\mathrm{s^{-1}\,)}\\

            \hline
            \noalign{\smallskip}

 44 & 11\ 42\ 20.25 ,+58\ 28\ 55.6 &  21.19 & 95157 &127   \\  
 45 & 11\ 42\ 20.29 ,+58\ 31\ 58.6 &  20.13 & 95813 & 81   \\  
 46 & 11\ 42\ 20.58 ,+58\ 29\ 08.4 &  18.15 & 96208 &132   \\  
 47 & 11\ 42\ 20.93 ,+58\ 29\ 43.5 &  19.00 & 97843 & 64   \\  
 48 & 11\ 42\ 21.35 ,+58\ 31\ 21.2 &  19.09 & 96142 & 48   \\  
\textit{49} & 11\ 42\ 21.57 ,+58\ 36\ 02.2 &  18.85 & 60383 & 62   \\  
 50 & 11\ 42\ 21.73 ,+58\ 33\ 55.4 &  19.56 & 96492 & 45   \\  
 51 & 11\ 42\ 21.79 ,+58\ 32\ 12.7 &  19.57 & 97021 & 68   \\  
 52 & 11\ 42\ 21.80 ,+58\ 33\ 10.6 &  19.08 & 99953 & 64   \\  
\textit{53} & 11\ 42\ 22.64 ,+58\ 33\ 33.6 &  19.54 & 46718 & 86   \\  
 54 & 11\ 42\ 22.66 ,+58\ 31\ 53.4 &  19.96 & 97289 & 51   \\  
 55 & 11\ 42\ 23.03 ,+58\ 30\ 44.2 &  18.17 & 97054 & 55   \\  
 56 & 11\ 42\ 23.16 ,+58\ 30\ 54.6 &  18.18 & 98169 & 53   \\  
 57 & 11\ 42\ 23.27 ,+58\ 31\ 26.2 &  19.00 & 96208 & 88   \\  
 58 & 11\ 42\ 23.40 ,+58\ 30\ 22.8 &  19.86 & 98791 & 90   \\  
 59 & 11\ 42\ 23.40 ,+58\ 31\ 02.1 &  20.60 & 98118 & 77   \\  
 60 & 11\ 42\ 23.40 ,+58\ 31\ 56.4 &  20.66 & 96609 & 92   \\  
 61 & 11\ 42\ 23.57 ,+58\ 31\ 13.1 &  19.50 & 94118 & 66   \\  
 62 & 11\ 42\ 23.97 ,+58\ 33\ 38.0 &  19.10 & 96433 & 64   \\  
 63 & 11\ 42\ 24.39 ,+58\ 33\ 43.6 &  20.20 & 98597 & 66   \\  
 64 & 11\ 42\ 24.62 ,+58\ 27\ 56.3 &  19.84 & 97922 & 54   \\  
 65 & 11\ 42\ 24.77 ,+58\ 31\ 24.9 &  19.24 & 93425 & 53   \\  
\textbf{66} & 11\ 42\ 24.78 ,+58\ 32\ 05.3 &  17.01 & 96531 & 28   \\  
\textit{67} & 11\ 42\ 25.34 ,+58\ 31\ 19.3 &  19.70 & 39316 &101   \\  
 68 & 11\ 42\ 25.37 ,+58\ 35\ 16.1 &  20.32 & 99142 &117   \\  
\textit{69} & 11\ 42\ 25.38 ,+58\ 31\ 18.9 &  19.00 & 39337 & 79   \\  
 70 & 11\ 42\ 25.44 ,+58\ 32\ 28.4 &  20.87 & 98894 & 84   \\  
 71 & 11\ 42\ 25.50 ,+58\ 31\ 57.6 &  19.46 &100922 &171   \\  
 72 & 11\ 42\ 25.68 ,+58\ 33\ 51.9 &  19.99 & 98377 & 86   \\  
 73 & 11\ 42\ 25.87 ,+58\ 28\ 53.2 &  19.66 & 95012 & 68   \\  
\textbf{74} & 11\ 42\ 26.25 ,+58\ 32\ 43.3 &  17.73 &100224 & 55   \\  
 75 & 11\ 42\ 26.42 ,+58\ 30\ 07.0 &  20.96 & 96765 &101   \\  
 76 & 11\ 42\ 26.83 ,+58\ 30\ 44.7 &  18.21 & 95862 & 53   \\  
 77 & 11\ 42\ 26.96 ,+58\ 34\ 19.2 &  19.85 & 98930 & 81   \\  
 78 & 11\ 42\ 27.09 ,+58\ 30\ 42.9 &  20.46 & 95532 & 88   \\  
 79 & 11\ 42\ 27.78 ,+58\ 30\ 32.3 &  18.22 & 96321 & 32   \\  
 80 & 11\ 42\ 27.80 ,+58\ 31\ 58.3 &  18.64 & 99455 & 30   \\  
 81 & 11\ 42\ 27.82 ,+58\ 29\ 01.4 &  19.48 & 95521 & 70   \\  
 82 & 11\ 42\ 28.29 ,+58\ 34\ 05.3 &  21.29 & 95519 &119   \\  
 83 & 11\ 42\ 28.73 ,+58\ 32\ 20.9 &  19.56 & 97331 & 77   \\  
\textit{84} & 11\ 42\ 29.46 ,+58\ 32\ 49.4 &  20.73 & 86542 & 77   \\  
 85 & 11\ 42\ 29.72 ,+58\ 32\ 04.3 &  20.07 & 99727 & 84   \\  
 86 & 11\ 42\ 30.20 ,+58\ 29\ 36.8 &  20.26 & 94471 & 59   \\  

            \noalign{\smallskip}			    
            \hline					    
            \noalign{\smallskip}			    
            \hline					    
         \end{array}
     $$ 
         \end{table}
\addtocounter{table}{-1}
\begin{table}
          \caption[ ]{Continued.}
     $$ 
           \begin{array}{r c r r r}
            \hline
            \noalign{\smallskip}
            \hline
            \noalign{\smallskip}

\mathrm{ID} &\alpha,\delta\,(\mathrm{J}2000) & r^{\prime} & v\,\,\,\,\,& \Delta v\\
 &                 & &\mathrm{(\,km}&\mathrm{s^{-1}\,)}\\

            \hline
            \noalign{\smallskip}
   
 87 & 11\ 42\ 30.21 ,+58\ 28\ 57.2 &  19.01 & 96268 & 53  \\  
\textit{88} & 11\ 42\ 30.35 ,+58\ 32\ 16.9 &  20.58 &132539 & 84  \\  
\textit{89} & 11\ 42\ 30.42 ,+58\ 32\ 51.7 &  20.51 &102690 & 68  \\  
 90 & 11\ 42\ 30.58 ,+58\ 29\ 49.3 &  20.14 & 95192 & 64  \\  
\textit{91} & 11\ 42\ 30.81 ,+58\ 32\ 55.8 &  21.33 & 87000 &220  \\  
 92 & 11\ 42\ 31.00 ,+58\ 30\ 50.5 &  19.94 & 97625 & 42  \\  
 93 & 11\ 42\ 31.14 ,+58\ 32\ 24.5 &  19.97 & 95732 & 53  \\  
\textit{94} & 11\ 42\ 31.61 ,+58\ 31\ 58.8 &  19.47 & 46323 & 66  \\  
\textit{95} & 11\ 42\ 31.65 ,+58\ 33\ 04.5 &  19.13 &101910 & 64  \\  
\textit{96} & 11\ 42\ 31.66 ,+58\ 33\ 10.1 &  20.31 & 39311 & 75  \\  
 97 & 11\ 42\ 31.71 ,+58\ 31\ 31.2 &  19.62 & 98490 & 73  \\  
\textit{98} & 11\ 42\ 31.80 ,+58\ 31\ 56.5 &  20.14 &146049 & 81  \\  
 99 & 11\ 42\ 32.00 ,+58\ 35\ 16.4 &  20.69 & 96579 &106  \\  
100 & 11\ 42\ 32.40 ,+58\ 29\ 51.6 &  20.66 & 93838 & 99  \\  
\textit{101} & 11\ 42\ 32.67 ,+58\ 31\ 48.1 &  19.46 & 46519 & 79  \\  
102 & 11\ 42\ 32.73 ,+58\ 32\ 17.0 &  20.58 & 96760 & 77  \\  
\textit{103} & 11\ 42\ 32.90 ,+58\ 31\ 53.6 &  20.29 &145458 & 86  \\  
104 & 11\ 42\ 33.02 ,+58\ 30\ 23.5 &  20.59 & 96943 & 64  \\  
105 & 11\ 42\ 33.11 ,+58\ 31\ 36.7 &  20.84 & 94982 & 86  \\  
\textit{106} & 11\ 42\ 33.37 ,+58\ 33\ 06.4 &  18.43 & 19387 & 55  \\  
107 & 11\ 42\ 33.77 ,+58\ 32\ 06.7 &  20.40 & 99125 & 88  \\  
108 & 11\ 42\ 34.12 ,+58\ 35\ 31.2 &  20.49 &100946 & 73  \\  
109 & 11\ 42\ 34.69 ,+58\ 34\ 14.4 &  21.16 & 99974 &101  \\  
110 & 11\ 42\ 35.38 ,+58\ 33\ 52.1 &  18.41 & 99892 & 35  \\  
\textit{111} & 11\ 42\ 35.46 ,+58\ 36\ 25.1 &  19.80 & 18306 &198  \\  
112 & 11\ 42\ 35.66 ,+58\ 28\ 48.1 &  19.56 &100490 & 51  \\  
113 & 11\ 42\ 37.71 ,+58\ 33\ 19.8 &  19.91 & 96834 & 44  \\  
114 & 11\ 42\ 38.92 ,+58\ 30\ 18.8 &  18.37 & 93762 & 51  \\  
115 & 11\ 42\ 39.92 ,+58\ 31\ 48.6 &  19.65 &100309 &112  \\  
\textit{116} & 11\ 42\ 41.88 ,+58\ 29\ 16.3 &  20.68 &143417 & 73  \\  
\textit{117} & 11\ 42\ 41.98 ,+58\ 29\ 52.9 &  20.65 &143144 & 97  \\  
118 & 11\ 42\ 42.99 ,+58\ 31\ 54.6 &  20.82 & 93791 &103  \\  
\textit{119} & 11\ 42\ 43.52 ,+58\ 28\ 32.4 &  20.99 &173827 & 75  \\  
\textit{120} & 11\ 42\ 43.64 ,+58\ 33\ 32.3 &  20.24 &112154 &158  \\  
121 & 11\ 42\ 44.29 ,+58\ 31\ 57.7 &  18.38 & 95717 & 48  \\  
\textit{122} & 11\ 42\ 44.65 ,+58\ 32\ 28.3 &  19.59 & 19164 &110  \\  
123 & 11\ 42\ 44.74 ,+58\ 34\ 04.8 &  19.62 & 96833 & 73  \\  
124 & 11\ 42\ 45.68 ,+58\ 35\ 17.0 &  20.86 & 99277 & 75  \\  
\textit{125} & 11\ 42\ 46.22 ,+58\ 31\ 13.4 &  18.98 &103511 & 53  \\  
126 & 11\ 42\ 46.51 ,+58\ 31\ 04.3 &  20.44 & 93695 & 88  \\  
\textit{127} & 11\ 42\ 48.29 ,+58\ 32\ 43.8 &  19.96 & 19592 & 48  \\  
\textit{128} & 11\ 42\ 48.48 ,+58\ 29\ 45.4 &  19.71 &110746 & 90  \\  
129 & 11\ 42\ 49.75 ,+58\ 34\ 15.6 &  19.29 & 98495 & 59  \\  
\textit{130}   & 11\ 42\ 51.26 ,+58\ 34\ 17.3 & 19.61 & 19534 &101 \\ 
131   & 11\ 42\ 51.82 ,+58\ 33\ 46.3 & 20.68 & 99361 & 66 \\ 
132   & 11\ 42\ 52.52 ,+58\ 32\ 54.3 & 20.90 & 99326 &108 \\ 
\textit{133}   & 11\ 42\ 53.87 ,+58\ 30\ 10.7 & 20.56 &124723 & 95 \\ 
\textit{134}   & 11\ 42\ 53.88 ,+58\ 30\ 55.8 & 19.94 & 74514 & 90 \\ 
\textit{135}   & 11\ 42\ 57.15 ,+58\ 33\ 10.9 & 20.52 &130786 &103 \\ 

            \noalign{\smallskip}			    
            \hline					    
            \noalign{\smallskip}			    
            \hline					    
         \end{array}
$$
\end{table}
